\documentstyle[12pt]{article}  
\begin{document}    
\def\o{\over}
\def\Ar{\rightarrow}
\def\bar{\overline}
\def\r{\gamma}
\def\d{\delta}
\def\a{\alpha}
\def\b{\beta}
\def\n{\nu}
\def\m{\mu}
\def\k{\kappa}
\def\e{\epsilon}
\def\p{\pi}
\def\th{\theta}
\def\om{\omega}
\def\vp{{\varphi}}
\def\Re{{\rm Re}}
\def\Im{{\rm Im}}
\def\ra{\rightarrow}
\def\t{\tilde}
\def\bar{\overline}
\def\l{\lambda}
\def\G{{\rm GeV}}
\def\M{{\rm MeV}}
\def\eV{{\rm eV}}

\setcounter{page}{1}
\thispagestyle{empty}
\begin{flushright}    
{EHU-98-08, July 1998}\\
\end{flushright}   
\vskip 1 cm
\centerline{\LARGE \bf Large Neutrino Flavor Mixings}
\vskip 0.5 cm
\centerline{\LARGE\bf  and Lepton Mass Matrices
\footnote{To appear in the proceedings of  the 1998
"INTERNATIONAL  SYMPOSIUM  on  LEPTON  and  BARYON  NUMBER  VIOLATION" at Trento (Italy).}}
\vskip 2 cm
\centerline{{\large \bf Morimitsu TANIMOTO}
  \footnote{E-mail address: tanimoto@edserv.ed.ehime-u.ac.jp} }
\vskip 1 cm
 \centerline{ \it{Science Education Laboratory, Ehime University, 
 790-8577 Matsuyama, JAPAN}}
\vskip 3 cm
\centerline{\large \bf Abstract}
\vskip 1 cm
 
 Recent atmospheric neutrino data at Super-Kamiokande suggest the near-maximal
 flavor mixing.
  Models for the lepton  mass matrix, which give the near-maximal flavor mixing,
  are discussed in this report.
   Mass matrix models are classified according to the mechanism  providing the large
    flavor mixing, and those are reviewed briefly.  "Naturalness" of the mass matrix is
	  also discussed in order to
     select the neutrino mass matrix. Details of the mass matrix with the $S_3$
	 flavor symmetry  are presented.
 \newpage
\section{Introduction}

The standard model (SM)   has still many unexplained features.     
It is a remarkable property of the quark and the lepton mass spectra that 
the masses of successive particles  increase by large factors.     
Mixings of the quark sector (CKM matrix) \cite{KM} seem also to have an hierarchical structure.      
Those features may provide an important basis for a new physics beyond the SM.                  

 On the other hand, the flavor mixing of the lepton sector 
 is very ambiguous. However, neutrino flavor oscillations  provide 
  information of the fundamental  property of neutrinos  such as  masses, 
  flavor mixings and  $CP$ violating phase.  
  In these years, there is growing  experimental evidences of the neutrino oscillations.
  The exciting one is the atmospheric neutrino deficit \cite{Atm1}$\sim$\cite{Atm3}
   and the solar neutrino deficit \cite{solar}.
	   Super-Kamiokande \cite{SKam} also presented the large
neutrino flavor oscillation in  atmospheric neutrinos.
 Furthermore, a new stage is represented by the long baseline(LBL) neutrino oscillation
  experiments.
   The first LBL  reactor experiment CHOOZ has already
  reported a  bound  of the neutrino oscillation \cite{CHOOZ},
    which gives a strong constraint of the flavor mixing pattern.
 The LBL accelerator  experiment K2K \cite{K2K} is planned to begin taking data
 in 1999,  whereas the MINOS \cite{MINOS} and ICARUS \cite{ICARUS}
  experiments will start in the first year of the next century.
  Those LBL experiments will clarify  masses,
  flavor mixings and $CP$ violation of neutrinos. 

The short baseline experiments may be helpful to understand  neutrino masses and
 flavor mixings.
	The tentative indication has been already given by the LSND experiment \cite{LSND},
   which is an accelerator experiment for $\n_\m\Ar\n_e(\bar \n_\m\Ar \bar \n_e)$.
   The  CHORUS and NOMAD experiments \cite{CHONOM}
   have  reported the  new bound for $\n_\m\Ar\n_\tau$ oscillation,
   which has already improved the E531 result \cite{E531}.
   The  KARMEN experiment \cite{KARMEN}
 is also searching for the $\n_\m\Ar\n_e(\bar \n_\m\Ar \bar \n_e)$ oscillation 
 as well as LSND. 
 The Bugey \cite{Bugey} and Krasnoyarsk \cite{Kras} reactor experiments and 
	  CDHS \cite{CDHS} and CCFR \cite{CCFR} accelerator experiments have already given
	bounds for the neutrino mixing parameters as well as E776 \cite{E776}.

  In this report,  our starting point as to neutrino mixings is
   the large $\nu_\m \Ar \nu_\tau$ oscillation of the atmospheric neutrino oscillation with 
 \begin{equation}
  \Delta m^2_{\rm atm}=  10^{-3}\sim 10^{-2} \eV^2 \ , \qquad\quad
 \sin^2 2\th_{\rm atm} \geq 0.8 \ ,
  \label{atm}
 \end{equation}
 \noindent
  which are derived from the recent data of the atmospheric neutrino deficit at Super-Kamiokande
   \cite{SKam}. 
 Then,  questions are raised:  Why is there large flavour mixing in the lepton sector
  in contrast with the quark sector?
  Are there possible mechanisms providing a large mixing angle
  from the lepton mass matrices, which are consistent with the quark sector?
  Answer is "Yes".  There are many mass matrix models to predict the  near-maximal
  mixing.
  


 \section{Origin of Near-Maximal Mixing} 
 
 In this section, we review lepton mass matrix models, which predict the near-maximal
 mixing between flavors.  We classify mass matrix models according to the mechanism
  giving the large mixing.
 \vskip 0.3 cm
 
 {\bf A:}\quad  See-saw enhancement
 
 The see-saw mechanism of neutrino mass generation gives a very natural and
 elegant understanding for the smallness of neutrino masses \cite{seesaw}.
 This mechanism may play another important role, which is to reproduce
  the large flavor mixing.
  In the standpoint of the quark-lepton unification, the Dirac mass
  matrix of neutrinos is similar to the  quark mass matrices.
  Therefore, the neutrino mixings turn out to be typically of the same order of 
  magnitude as the quark mixings.  However, the large flavor mixings of neutrinos could be
  obtained in the see-saw mechanism as a consequence of a certain structure of the
  right-handed Majorana mass matrix \cite{en1}\cite{en2}.  
  That is the so called see-saw enhancement of the neutrino
  mixing due to the cooperation between the Dirac and Majorana mass matrices.

  Mass matrix of light Majorana neutrinos $m_\n$ has the following form
  \begin{equation}          
   m_\n \simeq -m_D M_R^{-1} m_D^T \ ,      
  \end{equation}          
\noindent  
where $m_D$ is the neutrino Dirac mass matrix and $M_R$ is the Majorana mass matrix
of the right-handed neutrino components.
Then, the lepton mixing matrix is \cite{en1}  
  \begin{equation}          
   V_\ell = S_\ell^\dagger \cdot S_\n \cdot V_s \ ,      
  \end{equation}    
  \noindent where $S_\ell$, $S_\n$ are transformations which diagonalize the Dirac mass
  matrices of charged leptons and neutrinos, respectively.
  The $V_s$ specifies the effect of the see-saw mechanism, i.e. the effects of
  the right-handed Majorana mass matrix.  It is determined by
   \begin{equation}          
   V_s^T m_{ss}V_s = diag(m_1,m_2,m_3) \ ,      
  \end{equation}  
  \noindent where
   \begin{equation}          
       m_{ss} = -m_D^{diag} M_R^{-1} m_D^{diag} \ .     
  \end{equation} 
  \noindent Here $m_i(i=1,2,3)$ are the masses of light neutrinos and
      \begin{equation}          
       m_D^{diag} \equiv diag(m_{1D},m_{2D},m_{3D}) \ ,     
  \end{equation} 
  \noindent is the diagonalized Dirac mass matrix of neutrinos.
  In the case of two generations,  the mixing matrix $V_s$  is easily investigated
   in terms of one angle $\th_s$ as follows:
  \begin{equation}          
       \tan 2\th_s = \frac{\sin 2\th_M \e_D(1-\e)}{\e-\e^2_D+\sin^2\th_M(1-\e)(1+\e^2_D)} \ ,     
  \end{equation} 
  \noindent 
  \begin{equation}          
       \sin^2\th_M= \frac{1}{(1-\e)(1-\e^2_D)} 
	   \left [\pm\left(1+\frac{m_2}{m_3}\right )\sqrt{\e_0\e}-\e-\e_D^2 \right ]\ ,     
  \end{equation} 
  \noindent 
  where
  \begin{equation}          
      \e\equiv \frac{M_2}{M_3} \ , \quad \e_D\equiv \frac{m_{2D}}{m_{3D}} \ ,
	  \quad \e_0\equiv  \frac{m^2_{2D}m_3}{m^2_{3D}m_2}  \ .
  \end{equation} 
  \noindent 
  Thus, the mixing angle is given in terms of the diagonal components in
  mass matrices.  In the range $4\e^4_D/\e_0 \ll \e \ll \e_0$ with
  $m_2 \ll m_3$, the mixing can be approximately by \cite{en1}
   \begin{equation}          
       \sin^2\th_s \simeq  \frac{\e^2_D}{\sqrt{\e_0\e}} \ .     
  \end{equation} 
  \noindent 
   The mixing becomes maximal value $\sin^2\th_s=1$  at $\e=4\e^4_D/\e_0$.
   That is the enhancement due to the see-saw mechanism.
   The rich structure of right-handed  Majorana mass matrix can lead to
    the maximal flavor mixing of neutrinos.
   
   In this estimate, mass matrices are assumed to be real.
   However, the Majorana mass matrix has a non-trivial phase even in the two generation
   model.  The see-saw enhancement condition should be modified including
   $CP$ violating phases \cite{en2}.  It was found that the see-saw enhancement could be obtained
   due to the phase even if the Majorana mass matrix is proportional to the unit matrix.
   
   Models which satisfy the see-saw enhancement were proposed on the early stage by
   Harvey, Reiss and Ramond \cite{Harvey}, and Babu and Shafi  \cite{Shafi} 
   in the framework of $SO(10)$. 
    Recent works based on the flavor $U(1)$ symmetry \cite{U1} are attractive examples 
	for the see-saw enhancement.
	The $U(2)$ symmetry \cite{Bar} was also studied focusing on the see-saw enhancement \cite{U2}.
	A successful mass matrix is also presented in the  phenomenological point of view \cite{Bando}.
	
	 \vskip 0.3 cm
	{\bf B:} \quad Type II see-saw model 
	
	The conventional see-saw mechanism for neutrino masses is implemented in gauge model such as
	$SO(10)$ or the left-right symmetric models. The general form of the see-saw mass matrix is
	\begin{equation}          
    \left (\matrix{ f v_L&  m_D \cr      
                 m_D^T &  f M_R \cr   }\right ) \ ,  
  \end{equation}          
\noindent  
which gives 
\begin{equation}          
   m_\n=  f v_L -m_D M_R^{-1} m_D^T \quad {\rm with} \quad v_L\simeq \lambda v^2_{EW}/v_R \ .
  \end{equation}          
\noindent  
This is called the type II see-saw formula \cite{MV}.
Recall that the conventional see-saw formula omits the first term.
If due to some symmetry reasons, $f_{ab}=f_0 \delta_{ab}$, then a degenarate neutrino
spectrum emerges.  In this model, the flavor symmetry $S_4$ or the horizontal $SU(2)_H$
guarantee the degenerate $f_{ab}$.
For instance, in the range of $v_R=10^{13}-10^{16} \G$, the desired value
 $v_L\simeq 0.01\sim 1 \eV$
 is quite reasonable. In the minimal SUSY $SO(10)$ model,
  the realistic neutrino mixings  $\sin^2 2\th_{\odot}\simeq 2.8\times 10^{-2}$ and
  $\sin^2 2\th_{\rm atm}\simeq 0.84$   have been obtained by putting
 experimental data of  masses and CKM matrix elements of the quark sector.
 
  \vskip 0.3 cm
{\bf C:} \quad Exotic fields     

The new particle may be essential for the large flavor mixing.
The Zee model is a typical model, in which charged gauge singlet Higgs boson
plays important role to give the maximal mixing \cite{Zee}.
Anti-GUT model also need  additional some  Higgs bosons \cite{FN}, which
lead to the large flavor mixing. 
 The mixing between the ordinary fermions  and  the exotic ones
  may be an origin of the large flavor mixing \cite{BB}\cite{Haba}.
  The exogenous mixing based on $SO(10)$ has made possible to investigate the
  flavor mixing quantitatively \cite{BB}.
  
  \vskip 0.3 cm
  {\bf D:}\quad  Large evolution of mixings by RGE's
 
 There may be another enhancement mechanism of the neutrino flavor mixing.
 The renormalization group equation (RGE) of the see-saw neutrino mass operators with
  dimension 5 has been investigated by some authors \cite{RGE1}$\sim$\cite{RGE3}.
  Babu, Leung and Pantaleone pointed out that the neutrino flavor mixing is enhanced 
  by the RGE in the MSSM under the special conditions of the mass matrices.
  The numerical analyses have been given in ref.\cite{RGE3} focusing on recent
  experimental data of the atmospheric neutrino deficit.
  
   \vskip 0.3 cm
   {\bf E:}\quad  Large mixing derived from the charged lepton mass matrix
  
  In the standpoint of the quark-lepton unification, the charged lepton mass matrix
  is considered to be similar to the down quark one.
  Then, the mixing following from the charged lepton mass matrix may be considered to be
   small in the hierarchical base of the quark mass matrix.  However, this  expectation
   is not true if the mass matrix is non-Hermitian.
   In the $SU(5)$ model, the left(right)-handed down quark mixings are related to
   the right(left)-handed charged lepton mixings because these fermions belong 
   to same representation $5^*$ such as
	 \begin{equation}          
   5^*:\  \psi_L= ( d^{c1},\ d^{c2},\ d^{c3},\ e^-,\ \n )_L \ , 
  \end{equation}          
\noindent  
 and the Yukawa couplings are given by $5^*_i 10_j 5^*_H$(i,j=1,2,3).
 This feature was nicely taken into consideration in models 
 of refs.\cite{SY} and \cite{Alb}.
 
 It should be noticed that  observed quark mass spectra and CKM matrix
 only constrains the down quark mass matrix as follows \cite{Ramond}:
 
 \begin{equation}         
   m_{\rm down} \sim \left (\matrix{    
                \l^4 & \l^3 & \l^3 \cr     
                    & \l^2 & \l^2 \cr    
                  &  & 1    \cr } \right )    \quad {\rm with} \quad \l=0.22 \ .
      \end{equation}        
\noindent    
   The three unknown entries are related to the left-handed lepton mixing
    in the $SU(5)$ model.
	Thefore, there is a room in the charged lepton mass matrix
	to provide a source of the large flavor mixing
	of the lepton sector.
	
  \vskip 0.3 cm
   {\bf F:} \quad Effective neutrino mass matrix
  
  The democratic mass matrix \cite{Demo} needs the large rotation in order to move to
  the diagonal base.  In the quark sector, this large rotation is canceled each other 
  between   down quarks and  up quarks.  
  However, the situation of the lepton sector is very different from
  the quark sector if the effective neutrino mass matrix $m_{LL}^\n$ is far from the democratic one.
   Details of the model is discussed in the section 4.
   \vskip 0.5 cm
  
  Since the result of the atmospheric neutrino at Super-Kamiokande
   excites the model building of the lepton mass matrix, new ideas and models
   will be presented in the near future.
   Our classification of  models with the large flavor mixing will be not enough.

 \section{Naturalness and Near-Maximal Mixing Angle}
 
 Since there are many models, which predict the large flavor mixing,
 the "naturalness" of the mass matrix is helpful to select models.
 The idea of the natural mass matrix was proposed in order to restrict     
severely the arbitrariness in the construction of the quark mass matrices \cite{Natu}.    
  
Let us consider the $2\times 2$ Hermitian quark mass matrix $M_i(i=u,\ d)$.     
Assume for simplicity that it can be diagonalized by some orthogonal matrices 
$O_i(i=u,\ d)$  as follows:     
\begin{equation}          
   O_i^T M_i O_i= M_i^{diag}\equiv \left (\matrix{ m_{i1} & 0 \cr      
                 0 & m_{i2} \cr   }\right ) \ , \qquad  (i=u,\ d ) \ ,      
  \end{equation}          
\noindent     
with         
\begin{equation}         
   O_i=\left (\matrix{    
                \cos \th_i & \sin\th_i \cr     
               -\sin \th_i & \cos \th_i \cr    
                      } \right ), \qquad           
       V_q=O_u^T O_d=\left (\matrix{    
                \cos \th_c & \sin\th_c \cr     
               -\sin \th_c & \cos \th_c \cr    
                      } \right ),        
\end{equation}        
\noindent        
where $\th_c=\th_d-\th_u$.     
Then the  mass matrices can be written as    
\begin{equation}          
    M_i=  \left (\matrix{ c_i^2 m_{i1}+  s_i^2 m_{i2} &  c_i s_i(m_{i2}- m_{i1}) \cr      
                 c_i s_i(m_{i2}- m_{i1}) & s_i^2 m_{i1}+  c_i^2 m_{i2} \cr   }\right ) \ ,     
				 \qquad  (i=u,\ d ) \ ,    
	\label{Matrix}      
  \end{equation}      
 \noindent    
 where $c_i\equiv \cos \th_i$ and $s_i\equiv\sin \th_i$.    
   The mixing matrix $V_q$ is invariant under the changes        
      \begin{equation}         
  O_d\Ar O O_d \ , \qquad\qquad    O_u\Ar O O_u \ ,      
  \label{Change}       
\end{equation}        
\noindent     
where $O$ is some arbitrary orthogonal matrix.        
Thus, using the fact that $\sin\th_c\ll 1$, we can assume both 
$\th_d\ll 1$ and $\th_u\ll 1$ without loss of generality.     
Taking into account the quark mass hierarchy, we set    
\begin{equation}          
  m_u= a\l^4  m_c \ , \qquad   m_d= b\l^2  m_s \ ,    
  \end{equation}     
  \noindent    
   where $a$ and $b$ are ${\cal O}(1)$ coefficients, and    
$\l\equiv \sin\th_c  \simeq 0.22$. 
Thus the mass matrices are expressed in terms of $\th_u$ and $\th_d$ as following:
\begin{equation}         
 M_u\simeq \left (\matrix{    
              a\l^4+\sin^2\th_u & \sin\th_u \cr     
              \sin \th_u & 1\cr    
                   } \right )m_c , \quad          
 M_d\simeq \left (\matrix{    
              b\l^2+\sin^2\th_d & \sin\th_d \cr     
              \sin \th_d & 1\cr    
                   } \right )m_s .        
\end{equation}        
There are three different options now for the angles $\th_u$ and $\th_d$:       
\begin{eqnarray}
  1.\ && \sin\th_d\sim \l \ , \quad \sin\th_u\sim \l   \ , \nonumber\\      
  2.\ && \sin\th_d\sim \l\ , \quad  \sin\th_u\leq \l^2  \ , \nonumber \\    
  3.\ && \sin\th_d\leq \l^2\ ,\quad  \sin\th_u\sim \l  \ . \nonumber      
\end{eqnarray}        
\noindent    
The options 1 and 3 are unnatural. 
Indeed, one would require a severe     
fine-tuning of the matrix element ${[M_u]}_{11}$ forcing     
${[M_u]}_{11}\simeq {([M_u]}_{12})^2$        
to arrive at the large $m_u/m_c\sim \l^4$ hierarchy.        
On the other hand, option 2 gives a natural quark mass matrix without fine-tuning.
Here the fine-tuning means a tuning of ${\cal O}(\l^2)$, which comes from the 
following  inspection.  
There is a  well known  phenomenological relation between the CKM  
matrix element and the quark mass ratio as  
\begin{equation}  
  |V_{us}|\simeq \sqrt{m_d\o m_s}\simeq \l \ .  
\end{equation}  
  \noindent  
The down quark mass ratio dominates  $V_{us}$, while the contribution of the up 
quark mass ratio is at most $\sqrt{m_u/ m_c}\simeq {\cal O}(\l^2)$.  
This  relation is consistent with the option 2.  
Thus, the criterion of ${\cal O}(\l^2)$  tuning is very useful in order to select 
the option 2 by excluding  options 1 and 3.  
The extension of the natural mass matrices to the three family case   
in quark sector is also given  in Ref.\cite{Natu}.    
\par    
Naturalness is also expanded to the lepton sector \cite{Masa}.   
If neutrino sector has the hierarchical mass structure, the naturalness argument 
is exactly the same as in the example just described for the quark sector.  
However,  in the neutrino sector  the inverse mass hierarchy for the flavor 
is still allowed by the constraints obtained in the disappearance experiments of 
the neutrino oscillations in Bugey \cite{Bugey}, Krasnoyarsk \cite{Kras}, 
CDHS \cite{CDHS} and CCFR \cite{CCFR}.     
Since the neutrino mixing is chosen to be $\sin\th_\n\sim 1$  in this case, one 
cannot always guarantee to have both $\th_\n\ll 1$ and $\th_E\ll 1$(charged 
lepton mixing) by the change in eq.(\ref{Change}).     
Therefore, we should reconsider  the naturalness of the quark mass  
matrices in the lepton sector.   
\par  
In the case of $s\simeq 1(c\ll 1)$ with $m_{1}\leq m_{2}$ in eq.(\ref{Matrix}), 
the mass matrix is expressed approximately:    
\begin{equation}          
    M_\nu\simeq  \left (\matrix{m_2 &  c m_2 \cr      
                 c m_2 & m_1+  c^2 m_2 \cr   }\right ) \ .     
\end{equation}     
The natural mass matrix without the fine-tuning requires     
\begin{equation}    
   m_1 \geq c^2 m_2 \ .    
\label{cond1}  
\end{equation}    
\noindent   
For example, if we take $c\simeq a'\l,\  m_1\simeq a\l m_2$, which satisfies the 
condition(\ref{cond1}), then matrix  
\begin{equation}         
 M_\nu\simeq \left (\matrix{    
               1 & a'\l \cr     
              a'\l & a\l \cr    
                   } \right )m_2   \ ,
\end{equation}  
gives us  the masses $m_2$ and $a\l m_2$, correspondingly.  
Thus, the naturalness condition follows from the (2,2) entry in the case of the 
inverse mass hierarchy of the flavor,  while in the quark sector this condition 
follows from the (1,1) entry.
\par    
It is necessary to clarify the concepts of naturalness for the large mixing 
angle.
Furthermore, the recent  experiments \cite{Atm1}$\sim$\cite{SKam} also indicate 
the large flavor mixing  in the neutrinos.
Let us consider the $2\times 2$ symmetric matrix $M_\nu$.
Again, assume for simplicity that it can be diagonalized by some orthogonal 
matrix $O_\nu$:    
\begin{equation}    
O_\nu^T M_\nu O_\nu=M_\nu^{diag}\equiv     
\left(\begin{array}{cc}    
 m_1 & 0 \\    
 0   & m_2 \\    
\end{array}\right).  
\end{equation}    
Here the orthogonal matrix $O_\nu$ is     
\begin{equation}    
O_\nu=\left(\begin{array}{cc}    
c & s \\    
-s & c \\    
\end{array}\right) \ , \qquad        
c\equiv\cos\theta_\nu\simeq\frac{1}{\sqrt{2}} \ , \qquad     
s\equiv\sin\theta_\nu\simeq\frac{1}{\sqrt{2}}.    
\end{equation}    
Then $M_\nu$ is expressed in terms of the mass eigenvalues and mixing  
angle as follows:     
\begin{equation}    
M_\nu=O_\nu M_\nu^{diag} O_\nu^T=\left(\begin{array}{cc}    
m_1c^2+m_2s^2 & cs(m_2-m_1) \\    
cs(m_2-m_1) & m_1s^2+m_2c^2 \\ \end{array}\right) .    
\end{equation}    
\par    
If the neutrino masses have the hierarchy such as $m_1/m_2 =\e \ll 1$ with     
the large mixing angle,     
 the mass matrix $M_\nu$ is    
\begin{equation}    
M_\n \simeq \left(\matrix{\e c^2+s^2  & cs \cr  cs    & c^2+\e s^2 \cr}  
\right)m_2 \ .    
\end{equation}    
In the (1,1) entry, $\epsilon c^2$ should be fine tuned against $s^2$ in order to 
arrive at $m_1/m_2\sim \e$.     
For example, taking $c\simeq s\simeq 1/\sqrt{2}$, the matrix  
\begin{equation}         
 M_\nu \simeq \left (\matrix{    
               \frac{1}{2} & \frac{1}{2}\cr     
             \frac{1}{2}  &  \frac{1}{2} \cr   
                   } \right )m_2  \ ,
\label{simple}
\end{equation}  
gives $m_1 \simeq 0$ and $m_2$.
Further if we assume that $\e\simeq a\l^2$, the matrix $M_\nu$ in (\ref{simple}) 
should be replaced by the following matrix
\begin{equation}         
M_\nu\simeq \left(\matrix{    
               \frac{1+a\l^2}{2} & \frac{1}{2}\cr     
             \frac{1}{2}  &  \frac{1+a\l^2}{2} \cr   
                   } \right )m_2  \ .  
\end{equation}  
The latter gives the mass $m_1=O(\l^2)m_2$ provided that the order of $O(\l^2)$ 
is fine-tuned.  
Thus, the hierarchical neutrino masses are unnatural in the case of the  
large mixing angle.     
\par    
In the case when neutrino masses are approximately degenerate such as    
$m_1 \simeq m_2$ and $(m_2-m_1)/m_2 =\e \ll 1$, the mass matrix $M_\n$ is    
\begin{equation}    
M_\n \simeq \left (\matrix{ 1  & cs\e \cr cs\e & 1 \cr} \right)m_2 \ .    
\label{nat1}    
\end{equation}    
\noindent 
Hence, no fine tuning is required for any entry.     
We can explore the same argument when     
the mass eigenvalues are $-m_1$ and $m_2$, in which case       
\begin{equation}    
M_\nu \simeq \left(\matrix{ c^2\e  & 2 cs \cr 2 cs  &  s^2\e \cr}  
\right)m_2 \ .\label{nat2}    
\end{equation}    
Therefore  in the case of the large  mixing angle we call     
eqs.(\ref{nat1},\ref{nat2})  the natural mass matrix.    
\par  
In the sequel the naturalness of the mass matrix for the lepton sector is 
investigated for the three family model.   
The  conclusion in the case of three families as follows:
 the neutrino mass matrices with quasi degenerate  masses and maximal mixing are the natural
   ones. Quasi degenerate masses means
   $m_3\simeq m_2 \gg m_1$ or  $m_3\simeq m_2 \simeq m_1$.

In the next section, we present a texture with  the $S_3$ symmetry, which is a
typical natural mass matrix.                 

\section{Neutrino Mass Matrix with $\bf S_3$ Symmetry}

One of the most attractive description of the quark sector in
the phenomenological mass matrix approach starts with an
$S_3(R)\times S_3(L)$ symmetric mass term (often called ``democratic''
mass matrix) \cite{Demo}:
\begin{equation}
M_q= {c_q \over 3}
             \left( \matrix{1 & 1 & 1 \cr
                            1 & 1 & 1 \cr
                            1 & 1 & 1 \cr
                                         } \right) \ ,
  \end{equation}
\noindent
where $q=$ up and down, and quarks belong to 
{\bf 3}={\bf 2}$\oplus${\bf 1} of $S_3(L)$ or $S_3(R)$. 
The same form is still available for the charged lepton sector.
However, the neutrino mass matrix is different if they are Majorana particles.
The $S_3(L)$ symmetric mass term is given as follows:
\begin{equation}
M_\n= {c_\n}  \left( \matrix{1 & 0 & 0 \cr
                            0 & 1 & 0 \cr
                            0 & 0 & 1 \cr  } \right)
	+ {c_\n} r \left( \matrix{0 & 1 & 1 \cr
                            1 & 0 & 1 \cr
                            1 & 1 & 0 \cr  } \right) \ , 
  \end{equation}
\noindent
where $r$ is an arbitrary parameter.
The  eigenvalues of this matrix are given as 
$c_\n(1+2r, \ 1-r, \ 1-r)$, which means that there are at least two degenerate masses
 in the $S_3(L)$ symmetric Majorana mass matrix \cite{FTY}\cite{KK}.
 
 If three degenerate  light neutrinos are required, the parameter $r$ should be
  taken as $r=0$ or $r=-2$.  The first case was discussed in ref.\cite{FTY}
   and the second case was discussed in ref.\cite{KK}.
   The difference of $r$ leads to the difference in the $CP$ property of  neutrinos.
   If $r=-1/2$, one finds two massive neutrinos and one massless neutrino.
   So the $S_3(L)$ symmmetry could be reconciled with the LSND data \cite{LSND}  
   by including the symmetry breaking terms.

   Alternative representation of the $S_3(L)$ symmetric mass matrix is given as
  \begin{equation}
  M_\n= {c_\n}  \left( \matrix{e^{i\a} & 1 & 1 \cr
                            1 & e^{i\a} & 1 \cr
                            1 & 1 & e^{i\a} \cr  } \right) \ ,
  \end{equation}
\noindent 
which is based on the universal strength for Yukawa couplings (USY) hypothesis \cite{Lisbon}.
If $\a=2\pi/3$, three neutrino masses are degenerate.

In order to reproduce the atmospheric neutrino deficit by the large neutrino
oscillation, the symmetry breaking terms are required.
Since  results are almost same, we show the numerical analyses
  in ref.\cite{FTY}, where the LSND data is disregarded.

 Let us start with  discussing the following  charged lepton mass matrix:
 \begin{equation}
M_\ell= {c_\ell \over 3}
             \left ( \matrix{1 & 1 & 1 \cr
                            1 & 1 & 1 \cr
                            1 & 1 & 1 \cr
                                         } \right )
+\left ( \matrix{\delta_1^\ell & 0 & 0 \cr
                         0 & \delta_2^\ell & 0 \cr
                         0 & 0 & \delta_3^\ell \cr
                                              } \right ) \ .
  \end{equation}
\noindent
 The first term is a unique representation of
the $S_3(R)\times S_3(L)$ symmetric matrix and the second one is a symmetry braking
matrix given by Koide \cite{Koide}.
This matrix is diagonalised as
\begin{equation}
U_\ell^\dagger M_\ell U_\ell = {\rm diag}(m_1^\ell, m_2^\ell, m_3^\ell) \ ,
\end{equation}
where
\begin{eqnarray}
m_1^\ell&=&(\delta_1^\ell+\delta_2^\ell+\delta_3^\ell)/3-\xi^\ell/6  \ ,\nonumber \\
m_2^\ell&=&(\delta_1^\ell+\delta_2^\ell+\delta_3^\ell)/3+\xi^\ell/6  \ , \\
m_3^\ell&=&c_\ell+(\delta_1^\ell+\delta_2^\ell+\delta_3^\ell)/3  \ ,\nonumber
\end{eqnarray}
with
\begin{equation}
\xi^\ell=[(2\delta_3^\ell-\delta_2^\ell-\delta_1^\ell)^2+3(\delta_2^\ell-\delta_1^\ell)^2
                                                ]^{1/2} \ . 
\end{equation}												

The matrix that diagonalises $U_\ell=AB_\ell$ reads
\begin{equation}
A= \left ( \matrix{1/\sqrt 2 & 1/\sqrt 6 & 1/\sqrt 3 \cr
                   -1/\sqrt 2 & 1/\sqrt 6 & 1/\sqrt 3 \cr
                           0 & -2/\sqrt 6 & 1/\sqrt 3 \cr
                                         } \right ) \ ,
 \end{equation}
\begin{equation}
B_\ell\simeq \left (\matrix{ \cos \theta^\ell & - \sin \theta^\ell & 
                                          \lambda^\ell\sin 2\theta^\ell \cr
                  \sin \theta^\ell  & \cos \theta^\ell & 
                                        \lambda^\ell\cos 2\theta^\ell   \cr
              -\lambda^\ell\sin 3\theta^\ell    & \lambda^\ell\cos 3\theta^\ell & 1 \cr
                                         } \right ) \ ,
 \end{equation}
with
\begin{equation}
\tan 2\theta^\ell\simeq-\sqrt{3}{\delta_2^\ell-\delta_1^\ell\over 
  2\delta_3^\ell-\delta_2^\ell-\delta_1^\ell} \ ,
 \quad \lambda_\ell={1 \over \sqrt 2}{1 \over 3c_\ell}\xi^\ell \ .
\end{equation}
 It has been shown 
 that all quark masses and mixing angles are successfully
given by taking $\delta_1=-\epsilon$, $\delta_2=\epsilon$ and
$\delta_3=\delta$.
 Analogous to the
quark sector, $\delta_1^\ell=-\epsilon_\ell$, $\delta_2^\ell=\epsilon_\ell$ and 
$\delta_3^\ell=\delta_\ell$ are taken.
 The three mass eigenvalues  are then 
\begin{equation}
m_1^\ell\simeq -\epsilon_\ell^2/2\delta_\ell, \hskip5mm m_2^\ell\simeq 2\delta_\ell/3+
\epsilon_\ell^2/2 \delta_\ell,
\hskip5mm m_3^\ell\simeq c_\ell+\delta_\ell/3 \ ,
\end{equation}
and the angle $\th^\ell$ is
\begin{equation}
\sin \theta^\ell \simeq - \sqrt{\left |{m_1^\ell\o m_2^\ell}\right |} \  . 
\end{equation}

Let us turn to the neutrino mass matrix:

 \begin{equation}
M_\n= {c_\n}
             \left ( \matrix{1 & 0 & 0 \cr
                            0 & 1 & 0 \cr
                            0 & 0 & 1 \cr
                                         } \right )
+\left (\matrix{0 & \epsilon_\nu & 0 \cr
                 \epsilon_\nu  & 0 & 0 \cr
                  0 & 0 & \delta_\nu \cr} \right ) \ ,
	\label{neumass}
  \end{equation}
 
  \noindent
 where the symmetry braking is given by  a small term with two adjustable parameters. 
An alternative natural choice to lift the mass degeneracy may be 
diag($-\epsilon_\nu, \epsilon_\nu, \delta_\nu)$, which we shall 
also discuss later.
The mass eigenvalues 
are $c_\nu\pm\epsilon_\nu$, and 
$c_\nu+\delta_\nu$, and the matrix
that diagonalises $M_\nu$  ($U^TM_\nu U=$diagonal) is

\begin{equation}
U_\nu= \left ( \matrix{1/\sqrt 2 & 1/\sqrt 2 & 0 \cr
                   -1/\sqrt 2 & 1/\sqrt 2 & 0 \cr
                           0 & 0 & 1 \cr
                                         } \right ) \ .
\end{equation}
 That is, our $M_\nu$ represents three degenerate neutrinos, with the degeneracy
lifted by a small parameters. 

The lepton mixing angle 
as defined by $V_\ell=(U_\ell)^\dagger U_\nu=(AB_\ell)^\dagger U_\nu$ is thus given
by

\begin{equation}
V_\ell\simeq
\left ( \matrix{1 & -(1/\sqrt 3)\sqrt{(m_e/m_\mu)} 
                                & (2/\sqrt 6) \sqrt{(m_e/m_\mu)} \cr
         \sqrt{(m_e/m_\mu)} & 1/\sqrt 3 & -2/\sqrt 6 \cr
                           0 & 2/\sqrt 6 & 1/\sqrt 3 \cr
                                         } \right ) \ , 
\end{equation}
where  the neutrino mass parameters  do not appear. 
The parameters $\epsilon_\nu c_\nu$ and $\delta_\nu c_\nu$ are fixed by the neutrino mass
difference explored by the oscillation effect.
 The normalisation $c_\nu$ is not
fixed unless one of the neutrino masses is known, but it is not important
for this argument, since 
the lepton mixing matrix is almost 
independent of the details of these parameters except for the 
$m_e/m_\mu$ ratio.
If we retain all small terms, the lepton mixing angle is predicted
to be

\begin{equation}
V_\ell= \left ( \matrix{0.998 & -0.045 & 0.05 \cr
                        0.066 & 0.613 & -0.787 \cr
                        0.005 & 0.789 & 0.614 \cr
                                         } \right ) \ ,
\end{equation}
\noindent which leads the large
$\nu_\mu-\nu_\tau$ oscillation $\sin^22\theta_{\rm atm}\simeq 8/9$.
 For the $\nu_e-\nu_\mu$ oscillation
$\sin^22\theta_\odot\simeq 8\times 10^{-3}$, which also agrees with the neutrino 
mixing corresponding to the
small angle solution of the MSW scenario \cite{MSW} for the solar neutrino problem \cite{BKS}.
It is remarked that predicted  $V_{\ell e3}\simeq 0.05$ is stable against the symmetry
breaking parameters.  In the future, this prediction will be tested in the
following  long baseline experiments $\n_\n \Ar \n_e$  and $\n_e \Ar \n_\tau$:
 \begin{eqnarray}
&& P(\n_\m\Ar \n_e) \simeq   4 V^2_{\m 3}V^2_{e3} \sin^2{\Delta m^2_{31} L \o 4 E} \ , \nonumber\\
&& P(\n_e\Ar \n_\tau) \simeq  4 V^2_{e3}V^2_{\tau 3} \sin^2{\Delta m^2_{31} L \o 4 E} \ .
\end{eqnarray}

A very important constraint comes from neutrinoless double beta decay
experiments. The latest result on the lifetime of 
$^{76}$Ge$\rightarrow$$^{76}$Se,
$\tau_{1/2}>1.1\times 10^{25}$ yr \cite{double} yields an upper limit on the Majorana
neutrino mass 0.4 eV \cite{Tomoda} to 1.1 eV \cite{Engel} depending on which nuclear model
is adopted for nuclear matrix elements.
We are then left with quite a narrow window for the neutrino mass 
$0.1 {\rm eV}\leq m_{\nu_e}\simeq m_{\nu_\mu}\simeq m_{\nu_\tau}\leq 
1 {\rm eV}$ for the present scenario to be viable.
It will be most interesting to push down the lower limit of neutrinoless
double beta decay lifetime. If the limit on neutrino mass is lowered by
one order of magnitude the our degenerate neutrino mass scenario 
 will be ruled out. 

The argument we have made above is of course by no means unique, 
and a different
assumption on the matrix leads to a different mass-mixing relation. 
Let us briefly discuss the
consequence of the other matrices we have encountered in the line of our 
argument above. If we adopt the symmetry breaking term alternative to eq.(\ref{neumass}), 
\begin{equation}
  \left ( \matrix{-\epsilon_\nu & 0 & 0 \cr
                 0  & \epsilon_\nu & 0 \cr
                  0 & 0 & \delta_\nu \cr} \right ) \ ,
\end{equation}
in parallel to the charged lepton and quark sectors,
we obtain the lepton mixing matrix to be
\begin{equation}
V_\ell\simeq \left ( \matrix{1/\sqrt 2 & -1/\sqrt 2 & 0 \cr
                   1/\sqrt 6 & 1/\sqrt 6 &  -2/\sqrt 6 \cr
                           1/\sqrt 3 &  1/\sqrt 3 &  1/\sqrt 3 \cr
                                         } \right ) \ .
 \end{equation}
\noindent
 This is identical to the matrix presented by
Fritzsch and Xing \cite{FX}.
  For this case one gets
\begin{equation}
\sin^22\theta_\odot\simeq 1,\hskip5mm \sin^22\theta_{\rm atm}\simeq 8/9 \ .
\end{equation}
\noindent
This case can accommodate the "just-so" scenario for the solar neutrino problem due to neutrino 
oscillation in vacuum \cite{BPW}.
This matrix has been investigated in detail \cite{Tanimoto} focusing on recent data
 at Super-Kamiokande \cite{Smir}$\sim$\cite{Goldhaber}.
 The "just so solution" leads to the bi-maximal flavor mixing, which 
  may be interesting for the  theoretical origin \cite{Barger}\cite{NY}.

 
\section{Summary}

  Atmospheric neutrino deficit at Super-Kamiokande suggests the near-maximal flavor mixing,
   which have excited the model building of the lepton mass matrices.
   Mass matrix models  are classified according to
   the mechanism  providing the large(maximal) flavor mixing.
   However,  more quantitative studies are needed in order to understand the
   origin of the large flavor mixings deeply.
   The studies of the lepton mass matrices will give clues of 
   new symmetry such as the flavor symmetry  and will indicate the particular directions
    for the unification of matter.
	Furthermore, the structute of the  neutrino mass matrix will give strong impact on
	other fields such as the leptogenesis \cite{FY}. 
	The $CP$ violating phase structure in the neutrino mass matrix is also a very attractive
	subject as well as the $CP$ violation of the quark sector.

\vskip 0.5 cm
{\large\bf Acknowledgements}

  I thank M. Fukugita and T. Yanagida for collaboration on the lepton mass matrix model
   with the  $S_3$ symmetry, 
   and M. Matsuda for the collaboration on the naturalness of the lepton mass matrices.
This research is  supported by the Grant-in-Aid for Science Research,
 Ministry of Education, Science and Culture, Japan(No.10140218, No.10640274).

\end{document}